\documentclass{article}

\PassOptionsToPackage{numbers, compress}{natbib}

\usepackage[preprint]{neurips_2025}

\usepackage[utf8]{inputenc} \usepackage[T1]{fontenc}    \usepackage{hyperref}       \usepackage{url}            \usepackage{booktabs}       \usepackage{amsfonts}       \usepackage{nicefrac}       \usepackage{microtype}      \usepackage{xcolor}         \usepackage{graphicx}         

\usepackage{algorithm}\usepackage{algorithmicx}\usepackage[noend]{algpseudocode}

\usepackage{caption}
\usepackage{makecell}

\title{LowKeyEMG: Electromyographic typing with a reduced keyset}

\author{Johannes Y. Lee$^{1*}$ \quad
Derek Xiao $^{1*}$ \quad 
Shreyas Kaasyap$^{1*}$ \quad 
Nima R. Hadidi$^{1}$ \quad \\
\textbf{John L. Zhou}$^{1}$ \quad
\textbf{Jacob Cunningham}$^{1}$ \quad 
\textbf{Rakshith R. Gore} $^{1}$ \quad 
\textbf{Deniz O. Eren}$^{1}$ \quad \\
\textbf{Jonathan C. Kao}$^{1}$ \quad 
\\
$^*$Equal contribution \\
$^1$University of California, Los Angeles \\
\texttt{\{johanneslee,derekxiao93\}@g.ucla.edu}\\
\texttt{\{skaasyap,nhadidi,johnzhou,jocunningham,rakshithrgore,deren\}@g.ucla.edu}\\
\texttt{kao@seas.ucla.edu}\\
}

\begin{document}

\maketitle

\begin{abstract}

We introduce LowKeyEMG, a real-time human-computer interface that enables efficient text entry using only 7 gesture classes decoded from surface electromyography (sEMG). 
    Prior work has attempted full-alphabet decoding from sEMG, but decoding large character sets remains unreliable, especially for individuals with motor impairments. 
Instead, LowKeyEMG reduces the English alphabet to 4 gesture keys, with 3 more for space and system interaction, to reliably translate simple one-handed gestures into text, leveraging the recurrent transformer-based language model RWKV for efficient computation. 
    In real-time experiments, participants achieved average one-handed keyboardless typing speeds of 23.3 words per minute with LowKeyEMG, and improved gesture efficiency by $17\%$ (relative to typed phrase length). When typing with only 7 keys, LowKeyEMG can achieve 99.2\% top-3 word accuracy, demonstrating that this low-key typing paradigm can maintain practical communication rates. 
    Our results have implications for assistive technologies and any interface where input bandwidth is constrained.
    \footnote{Videos: \url{https://johannes-lee.github.io/lowkeyemg}}
\end{abstract}

\section{Introduction}

In several settings, humans aim to achieve high-bandwidth communication with relatively low-bandwidth and low-dimensional inputs. 
For example, people with motor disability and disease, including amyotrophic lateral sclerosis, stroke, spinal cord injury (SCI), or amputation, lose naturalistic and fluid movements.
This has significant consequences on their communication and autonomy; for example, they may lose the ability to type on a keyboard, limiting their interaction with the digital world. 
One approach to restore movement and communication is to directly decode neural activity into control signals that can be used for typing and communication on a computer \cite{sivakumar2024emg2qwerty, ctrl2024generic, Willett2021-kg}.

In this work, we aim to decode typing via non-invasive sEMG activity, which reflects muscle and motor neuron activity.
Motor neuron activity may persist even as muscular coordination or activation is degraded, including in stroke \cite{olsen2023wrist, Meyers2024-ty} and SCI \cite{ting2019wearable, oliveira2024direct}. 
But this approach suffers from a fundamental signal information limitation issue: decoding more than $10$ classes from EMG activity consistently performs worse than decoding fewer classes \cite{olsen2023wrist, Meyers2024-ty, ting2019wearable, oliveira2024direct, Battraw2024-rl}.
As decoding accuracy decreases, resulting in more frequent errors, systems become less usable. 
For example, \citet{sivakumar2024emg2qwerty} found that EMG-based typing systems for healthy people were “usable” when decoding character error rate (CER) was approximately $10\%$ or less, yet some users were not able to reach this CER requirement.
This suggests that in settings where people are disabled and may therefore only be able to decode a relatively small number of classes reliably, or when healthy users cannot achieve reliable performance with many gestures \cite{sivakumar2024emg2qwerty}, we need to design communication systems that can translate reduced keysets (classes) into fluid typing.

We therefore pose the \textit{key} question: can we design an efficient and reliable typing system with a reduced keyset, and how small can we reduce the keyset to?

\begin{figure}
  \centering
  \includegraphics[width=5.5in, trim = 0cm 0.0cm 0cm 0.0cm, clip]{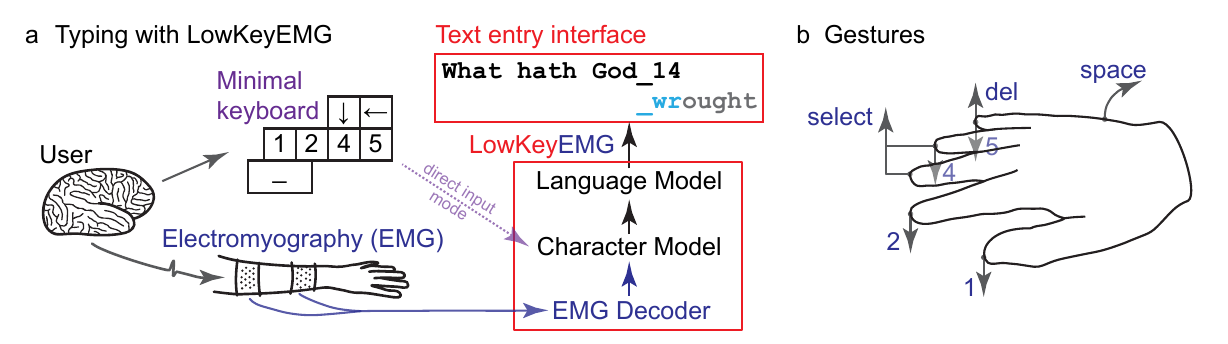}
  \caption{
  \textbf{LowKeyEMG text entry.}
  \textbf{(a)} Users can type with 7 or fewer keys using gestures (b) decoded from electromyography (EMG) or with other physical or virtual keyboards.
  \textbf{(b)} Gestures used during EMG typing. Alphabetic keys 1, 2, 4, 5: press thumb, index, ring, or little finger. Space: wrist ulnar deviation. Undo (``del''): lift little finger. Select: lift both middle and ring fingers.
  } \label{fig:overview}
\end{figure}

Leveraging recent advances of recurrent transformer-based large language models (LLMs) \cite{peng2023rwkv}, we design and demonstrate a real-time system (LowKeyEMG) that can reconstruct the typed word with $99.2\%$ top-3 accuracy when provided context, using only 4 alphabetic+punctuation keys, 1 space key, and 2 system interaction keys (select and undo), with each key corresponding to a gesture performed by a single hand (Figure \ref{fig:overview}).
We then test this system in real-time closed-loop experiments with healthy human participants, streaming and decoding EMG signals while running the LLM locally on a laptop (Intel i7-12800H, Nvidia RTX 3070 Ti Mobile, 8GB VRAM). Participants obtained day-average one-handed keyboardless sEMG typing speeds of $7.7$ to $19.6$~words per minute (wpm) without word completion, and $16.4$ to $27.8$~wpm with word completion and passage context, while always successfully typing each prompt (success rate: $100\%$).

Our results demonstrate the feasibility of reliable text entry with a substantially reduced keyset, with particular application for motor impairment where signal decoding is limited, thereby reducing the motor precision required of each user.
Certainly, when more than 7 keys can be decoded, additional keys can be used to represent the alphabet or be used as other control inputs, which may increase performance.
We further note that LowKey typing systems can be applied to any text entry interface where a reduced keyset is desirable, rather than only to EMG typing.
These may include eye gaze typing systems, physical or virtual keyboards, and switch-based systems such as with morse code or with binary codes, where each key is represented by a series of gestures.

\section{Related Work} \label{related_work}

There have been several efforts to decode typing with EMG signals based on QWERTY keyboard typing \cite{sivakumar2024emg2qwerty, crouch2021natural} and handwriting \cite{ctrl2024generic}.
\citet{crouch2021natural} decoded keyboard typing from 8 hours of individual data, and noted that the most common errors occur between characters that share a finger, indicating that cross-finger differentiation is relatively more robust.
CTRL-labs \cite{ctrl2024generic} focused on a generalizable interface that can be operated by na\"ive users, achieving an adjusted words per minute of 17.0~aWPM with handwriting.
CTRL-labs also achieved 93\% balanced decoding accuracy across 7 discrete gestures in the same study, a mode they did not explore for typing.

With the goal of reducing the number of keystrokes for motor-impaired individuals, Cai et al. \cite{caiUsingLargeLanguage2024, cai2022context, li2023c} developed an eye gaze typing interface with abbreviated sentence typing of the first character of each word.
Their system, however, required users to perform non-sequential actions to correct errors, and used a cloud-based transformer language model with latencies ranging from 240-660~ms \cite{cai2022context}, depending on the size of the language model.
Generally, eye gaze typing systems are fatiguing and require constant use of the eyes, precluding their use for other functions such as monitoring and correcting typed text. 

Previous systems for typing with fewer keys than alphabetical characters have been demonstrated and used in commercial applications.
For example, the T9 keyboard was used extensively in mobile phones, with 8 physical keys associated with up to 3-4 characters each, and additional keys for space, backspace, and scrolling through suggested words, totaling 11 keys.
The T9 keyboard fell out of favor as users moved toward touchscreen devices that achieve faster typing speeds. \section{Methods} \label{sec_methods}

\subsection{Recording}

We recorded EMG data at $2000$~Hz with 2 custom-made saline sponge electrode arrays (Ant Neuro) wrapped around a single forearm  (one array distal and one proximal) using the eego rt amplifier.
Each array had 32 electrodes, and both arrays shared external ground and reference electrodes that were attached to the same forearm.
Data were average referenced within each array and filtered with a 4th-order Butterworth bandpass filter from 10Hz to 999Hz and notch filters at 60, 60, 120, 180, 240, and 300~Hz with quality factors of 10, 4, 5, 2, 2, and 2, respectively.
Before each experiment, we visualized the filtered data to ensure that movement and disconnection artifacts were minimized.

\subsection{Experimental participants}
We recruited 3 healthy participants (ages 25-33) to perform EMG typing experiments.
All experiments were approved by the Institutional IRB.
We acquired informed consent from all participants, acknowledging the risk of skin irritation due to EMG recording.
Participants received a \$20 Amazon gift card for each hour of experimental participation.

\subsection{Gestures}
\begin{figure}
  \centering
\includegraphics[height=1.5in, trim = 0cm 0.0cm 0cm 0.0cm, clip]{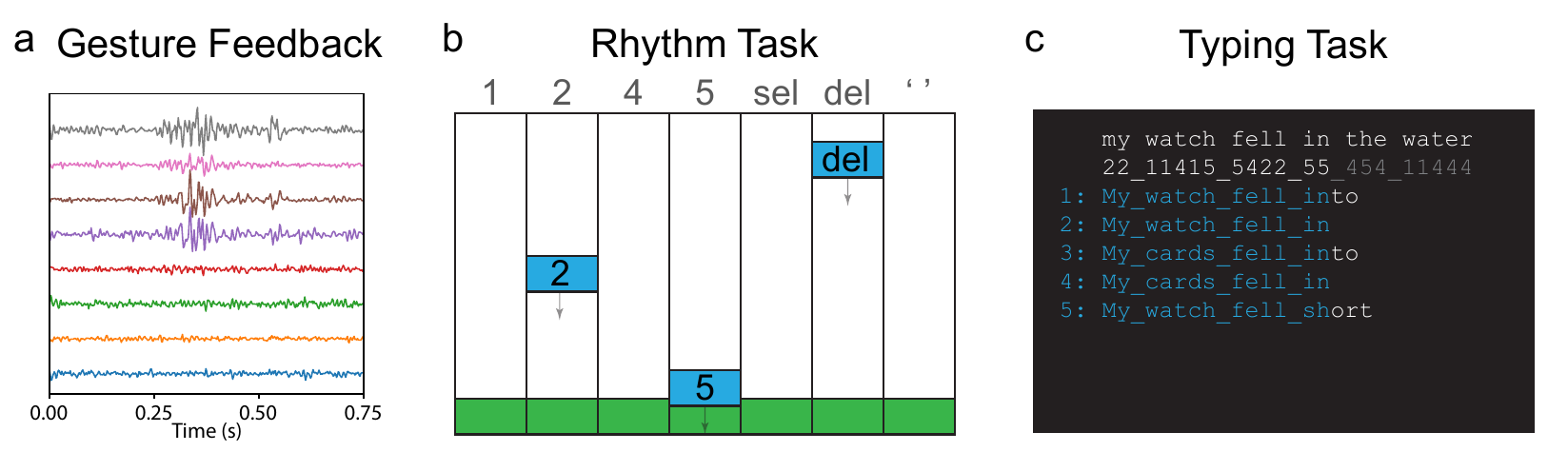}
  \caption{
  \textbf{Experimental tasks.}
  \textbf{(a)} Electromyography (EMG) signals were streamed live to help the subject develop intuition for gesture intensity and duration. 
  \textbf{(b)} After developing familiarity with the gestures, subjects played the rhythm game, where blocks in each column corresponds to 1 of 7 possible gestures. EMG activity during and between these gesture events were later used as training data for the gesture detector and classifier.
\textbf{(c)} These models were then deployed online to decode EMG signal on the fly, and allowed the user to type natural language passages using only 4 alphabetic classes, and 3 classes for \textit{select} (``sel''), \textit{undo} (``del''), and \textit{space}.
  }
\label{fig:methods}
\end{figure}

Participants were instructed to rest their right arm in a comfortable position with their elbow on an armrest. Their wrists were rested on a table with their palms face down and fingers on the table. 
They were instructed to prevent the electrodes from contacting the table or armrest. 
For alphabetical and punctuation characters, participants were instructed to press their thumb, index, ring, and little fingers into the table (Figure \ref{fig:overview}b).
They were also instructed to perform ulnar deviation of their wrist for \textit{space}, lift their little finger for \textit{undo}, and lift their middle and ring fingers for \textit{select}, which queued the next candidate for selection.
Participants were shown a livestream of recorded signals to help adjust the strength, duration, and posture of gestures (Figure \ref{fig:methods}a).
Participants were told to keep their gestures as short as they could.
During the typing task, users were encouraged to seek gestures that were classified well by the decoder, and to incorporate any changes for future sessions.

\subsection{Rhythm task: training data collection} \label{sec:rhythm}
To collect gesture data with aligned labels, participants performed a ``rhythm task'' at the beginning of each day, where users were prompted to perform one of the 7 gestures at specific time intervals (every 1-1.5~s), similar to the popular game Guitar Hero (Figure \ref{fig:methods}b).
On each day, we collected 80 events total per gesture, totaling 560 gesture events.
For each of 6 force and posture variations (light and firm presses, combined with a neutral wrist position, ulnar deviation, and radial deviation), we collected 9 events per gesture in a random order.
The first 448 gestures were used for training, the next 56 events were used for validation, and the last 56 events were withheld as a test set.
Participants were instructed to gesture naturally for the last 182 events.

\subsection{Gesture detection and classification}

\textbf{Gesture classifier and training}: Gesture decoding was composed of two stages: gesture detection and gesture classification.
Both stages were implemented with CNNs using the EEGNet architecture (Supplementary Table \ref{table:eegnet_hyper}) \cite{lawhern2018eegnet}. 
Inputs were \(64 \times 240\) (channels × time samples; 120 ms total duration).

These EMG activity windows containing gestures were realigned to be centered around the maximum absolute signal averaged across channels, within 150-300~ms of each prompted event.  
We split gestures from the rhythm task chronologically with an $80\%/10\%/10\%$ (448/56/56 gestures) training/validation/test split, and selected the model checkpoint with the highest test accuracy among the 5 model checkpoints with the highest validation accuracy. 
Checkpoints were saved every epoch. 
To augment the training data, we generated additional samples by shifting the window in 16 evenly spaced strides, such that the peak signal moved progressively from the start to the end of the window. 
Gaussian noise with a standard deviation equal to  $0.25$ times the channel-wise variance of the window was added to each augmented sample to improve robustness. 
During each epoch, we sampled augmented windows without replacement to bring the number of training samples from 448 to 5600. 

\textbf{Gesture detection}:
For gesture detection, we selected blank windows from the contiguous blocks of EMG activity between each gesture event, with anywhere from $150$ to $300$~ms of buffer around each gesture event, depending on the timing variance of the subject. 
We performed the same Gaussian noise dataset augmentation as in the gesture classifier.

\textbf{Real-time gesture classification}:
During online decoding, we performed gesture detection with $120$~ms non-overlapping sliding windows.
When a gesture was detected, the window centered around the time of maximum average amplitude over channels was used for gesture classification.
We limited gestures to at most once every 0.2~s or 0.4~s, depending on the participant, to reduce spuriously decoded gestures.
While a sequential decoder and the CTC loss may have improved performance compared to two-stage decoding, they were not necessary to achieve proficient one-handed typing speeds.

\subsection{LowKeyEMG typing task}
Participants used LowKeyEMG to type sentences by performing gestures outlined in Section \ref{sec:rhythm} (Figure \ref{fig:methods}c).
We compiled a list of 42 newly composed short passages (at least 4 sentences) to represent natural language outside of the language model's training corpus. 
This list included unpublished academic writing, text message conversations, informal writing, and unprompted writing for one paragraph. 
The first $40$ characters (without truncating any word) of the last sentence of each of these paragraphs were used as the prompt, with previous sentences as the context.
Participants were allowed to practice typing until they felt comfortable to proceed.
Participants were instructed to type each prompt as quickly as they could, and were allowed to take breaks in between trials.

We tested 3 features -- passage context, word completion, and Levenshtein distance-1 matching -- for their impact on user typing speed.
With passage context, the previous sentences of each passage are input to the LM prior to the prompt (which an n-gram LM would not be able to fully utilize). This models real-world text which is typically conditioned on previous text.
With word completion, LowKeyEMG uses the gesture subsequence spanning from the last space or selection up to the last character typed to identify candidate words that begin with that subsequence (i.e., prefix matches). 
Users could opt at any time to select a desired word completion among proposed candidates.
With Levenshtein distance-1 matching (d1), users could make mistakes, up to an edit distance of exactly 1, but still type 4+ letter words.
Words with an edit distance of 1 were applied a character model score penalty of $\mathrm{log}_{10}(0.01)$, rather than the $0.0$ score penalty for exact matches.
Participants typed each phrase in three conditions where passage context, word completion, and Levenshtein distance-1 matching were respectively: (A) off, off, off (``base''), (B) off, on, on (``completion''), and (C) on, on, on (``completion+context'').

The three conditions were randomly interleaved such that participants typed each phrase three times in a row with a random order of conditions.
While users were not directly told which condition was active for each trial, due to the differences in task feedback, users could guess which condition was active.
Because we suspected that users could not take advantage of all features of the typing system when conditions were interleaved, during a separate session, participants typed using only the completion+context condition to quantify the best typing performance. 

On the typing interface, participants were shown the prompted text at the top and a sequence of numbers and underscores (spaces) corresponding to the mapped sequence of gestures on the following line.
As participants typed the gesture sequence, the numbers would turn red if they were incorrect to reduce participants' cognitive load.
Candidates appeared below the displayed numbers, and the numbers were converted to text after candidates were selected and committed.
Character capitalization was not enforced.
For an example of a real-time experiment with LowKeyEMG, please see Supplementary Video 1.

\begin{algorithm}[t!]
\caption{LowKeyEMG beam search with tokenized LM}
\label{alg_lm_beam}
\begin{algorithmic}
\State $\mathcal{D} \gets $ dictionary of valid words (or tokens)
\State $\mathcal{G} \gets $ mapping of characters to gestures
\State $C_{\mathrm{prev}} \gets \{\textrm{(``'')}\}$: candidates of previous words, used to build $C$
\State $C \gets \{\textrm{(``'')}\}$: current candidates to be displayed
\State $s \gets 0$: selection index ($1$-indexed)
\For{Key $z_i$ in $\{z_1, ..., \mathrm{SPACE}, z_{L_1 + 2}, ..., \mathrm{SPACE}, ...\}$}
    \If{$z_i = \mathrm{BACKSPACE}$} 
        \State Load previous searcher state (i.e. undo)
        \State \textbf{continue}
    \EndIf
    \If{$z_i = \mathrm{SELECT}$}
        \State $s \gets s + 1$: Queue next candidate
        \State \textbf{continue}
    \EndIf
    \If{Selection is queued}
        \State $C_{\mathrm{prev}} \gets \{c_s\}, s \gets 0$: Commit queued selection of candidate with index $s$
        \State $\mathbf{S} \gets$ True: whether a selection was just committed
    \EndIf
\vspace{-6pt}
    \State \[ \hspace{-10pt} M \gets \left\{ \begin{array}{lll}
             \textrm{All valid tokens,} & \textrm{if } z_i = \textrm{SPACE } \& \textrm{ }\mathbf{S}\\
             \textrm{Exact matches and close matches,} & \textrm{if } z_i = \textrm{SPACE } \& \textrm{ not }\mathbf{S} \\
             \textrm{Exact matches, close matches, and prefix matches,}& \textrm{if } z_i = \textrm{CHAR}
        \end{array} \right.\] 
\State \textit{Perform beam search with beam width $K$:}
    \State $C_{\mathrm{search}} \gets \{(c_k, w_m) \textrm{ for all } c_k \textrm{ in } C_{\mathrm{prev}}, \textrm{words } w_m \textrm{ in } M\}$: candidates to search
    \vspace{1pt}
    \For{$j$ in $\mathrm{max}_M(\mathrm{token\,length\,of\,words})$}
        \For{$(c_k, w_m)$ in $C_{\mathrm{search}}$ with word $w_m = $ tokens $ t_1, t_2, ..., t_l$}
            \vspace{2pt}
            \State \textit{Use language model} LM \textit{and character model} CM \textit{to calculate}:
            \vspace{1pt}
            \State $\mathrm{Score}(c_k, w_m, j) = \mathrm{log}_{10} \left( \mathrm{LM}(t_j|c_k, t_1, ..., t_{j-1}) \cdot \mathrm{LM}(c_k, t_1, ..., t_{j-1}) \cdot \mathrm{CM}(w_m) \right)$
        \EndFor
        \State $C_{\mathrm{search}}  \gets $ sequences $\{(c_k, w_m)\}$ with top $K$ scores
    \EndFor
    \State $C \gets \{c_k + w_m \textrm{ for } (c_k, w_m) \textrm{ in } C_{\mathrm{search}}\}$. Display $C$.
    \vspace{2pt}
    \If{$z_i = \mathrm{SPACE}$}
        \State $C_{\mathrm{prev}} \gets C$: Store candidates as candidates of previous words
    \EndIf
\EndFor
\end{algorithmic}
\end{algorithm}

\subsection{LowKeyEMG search} 

To translate each key sequence to text, we adapt word beam search \cite{scheidl2018word} to a pre-trained tokenized language model (RWKV, \texttt{RWKV/rwkv-4-169m-pile} checkpoint on Hugging Face \cite{wolf2020transformers}).
We chose the transformer-based RWKV architecture because of its performance, its longer effective context length than traditional n-gram LMs, and because the computation time of next token predictions does not scale with context length at all due to its recurrent formulation \cite{peng2023rwkv}. 

We applied beam search to filter candidate suggestions for the user (Algorithm \ref{alg_lm_beam}), integrating a language model (LM), which scores next-token probabilities, and a character model (CM), which filters and scores probabilities of words based on their edit distance with the inputted key sequence.
The beam search candidates were updated after every detected keypress, taking $20 \pm 17$~ms to compute on average for each beam search step (Nvidia RTX 3070 Ti Mobile).
Users could queue a candidate for selection by using the \textit{select} gesture one or more times, committing the selection after decoding any non-\textit{undo} gesture.
Candidates were scored using the sum of their language model log probabilities (temperature=1.0) and their character model log probabilities.
Since we assumed users would be accustomed to typing spaces between every word, and because some multi-token words would be pruned prematurely when purely using Algorithm \ref{alg_open_lm_beam}, we used a closed dictionary of words (\texttt{american-english-large} from Linux distributions) and required spaces to be typed between words.
When \textit{space} was decoded after typing a word, only words that matched (either exactly, or with a Levenshtein distance of 1) the characters since the last \textit{space} or \textit{select} were considered.
We then cached the resulting candidates for combination with following words.
In order to accommodate multi-token words, we performed beam search at the token-level with candidates processed in parallel, keeping only the top $K=30$ candidates with the highest log probability.

\begin{algorithm}[t!]
\caption{Token-level Beam search with LM}
\label{alg_open_lm_beam}
\begin{algorithmic}
\For{Key $z_i$}
    \For{candidate $c \in C$}
        \State Score Previous token hypothesis: $z_i$ belongs to the previous token/word of $c$
        \State Score New token hypothesis: $z_i$ belongs to a new token/word of $c$
    \EndFor
    \State $C \gets $ Sort and keep top $K$ new candidates
\EndFor
\end{algorithmic}
\end{algorithm}

\subsection{Layout optimization}
To optimize the layout of characters vs. an alphabetically-arranged layout, we minimized the expected number of collisions between words, where words had the same gesture sequence, using words in the Brown corpus and their frequencies.
We optimized the probabilities (represented by logits) that each character would be assigned to each gesture.
We obtained optimized character-gesture layouts by iteratively sampling character-gesture assignment using the Gumbel softmax function and performing gradient descent on the number of word-word collisions, weighted by the sum of the frequencies of each pair of colliding words.
This resulted in the following key assignments: Key~1~=~\{a, c, d, u, v, w, x\}, Key~2~=~\{g, j, k, l, m, o, p, y\}, Key~4~=~\{b, e, r, t, z, comma\}, Key~5~=~\{f, h, i, n, q, s, period, question mark\} (Supplementary Table \ref{table:layouts}).

\section{Results} \label{sec_results}
\subsection{Offline gesture decoding}

Decoders were consistently trained to above $90\%$ test accuracy on same-day rhythm task data for both detection and classification components.
Despite this, decoding performance was typically less accurate during closed-loop typing task experiments, with higher rates of incorrect (undone) gesture emissions (Fig. \ref{fig:wpm}b). 
This is consistent with the brain-computer interface literature, which observes mismatch between \textit{offline} vs. \textit{online, real-time} decoder performance \cite{Chase2009-rk, Cunningham2011-pn, Fan2014-tu}.

\subsection{LM-enabled typing with few gestures}

\begin{figure}[t!]

    \centering
\includegraphics[width=5.5in, trim = 0cm 0.0cm 0cm 0.0cm, clip]{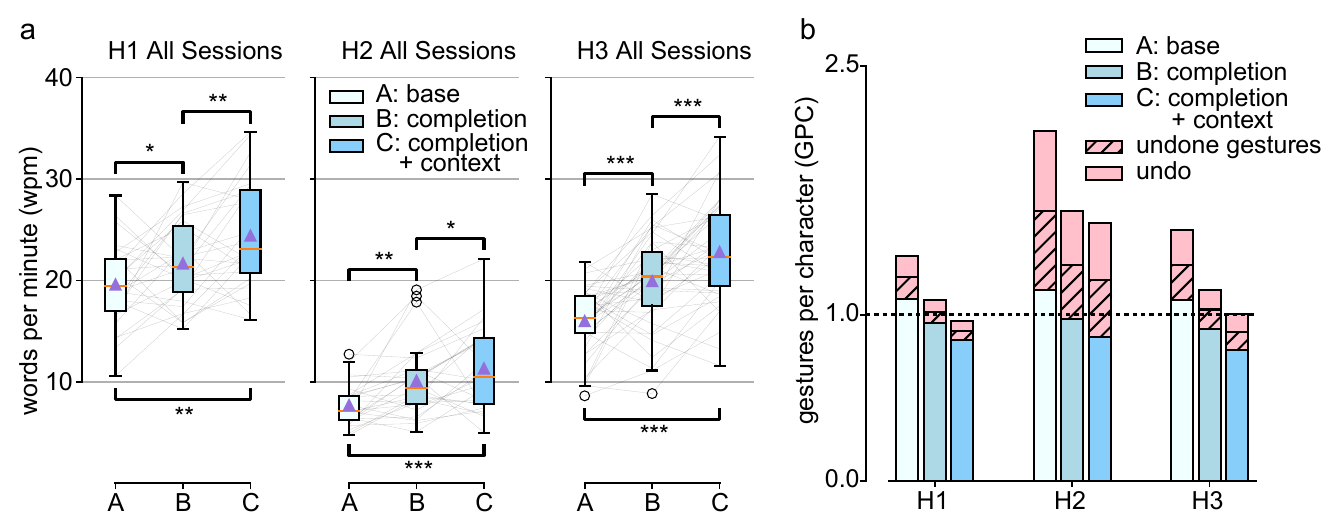}
    \caption{ 
    \textbf{Experimental performance.}
    \textbf{(a)} Typing speeds across three conditions for the 3 participants, A:~LowKeyEMG-base, B:~LowkeyEMG-completion, C:~LowKeyEMG-completion+context. 
    Lines connect the same phrase typed in each condition.
Purple triangle: mean, orange bar: median. One-sided Wilcoxon signed-rank test: $p<$~*$0.05$, **$0.01$, ***$0.001$.
    \textbf{(b)} Participants achieved reduced GPC (with and without error gestures) with word completion and context.  
Colors represent whether each gesture corresponded to a correct key (blue), incorrect key (shaded red), or undo of an incorrect key (solid red).
    }
    \label{fig:wpm}

\end{figure}

Using LowKeyEMG-base, without word completions or context, participants achieved an average typing speed of $14.4$ words per minute (wpm = 0.2*characters per minute). 
When using LowKeyEMG-completion, participants achieved an $11$-$25\%$ increase in typing speed, resulting in an average typing speed of $17.2$ wpm.
Finally, when using LowKeyEMG-completion+context, participants achieved a further $12$-$14\%$ increase over LowKeyEMG-completion. The final average typing speed with word completion and context was 19.5 wpm. 
Notably, the gains from adding word completion were larger for slower typists. The fastest typist, H1, noted that the cognitive load to look between the target gesture sequence and candidate suggestions discouraged frequent usage of word completions. For participant-specific typing speeds, please refer to Table \ref{table:typing_speed} and Figure \ref{fig:wpm}a. 
Together, these results indicate that participants can type at speeds approaching $20$~wpm with only 7 keys, and further, that incorporating word completion and context further increase performance (\href{https://johannes-lee.github.io/lowkeyemg/#SupplementaryVideo1}{Supplementary Video 1}).

\subsection{Per user maximum typing speeds}
We quantified how fast each participant would type if they developed more expertise. 
An expert user would more skillfully anticipate the suggestions, and also be able to type most words more automatically. 
We had participants perform a final session, where each user typed each phrase three times in succession using only LowKeyEMG-completion+context. 
This allowed participants to develop familiarity with the gesture sequence and the LM suggestions, and consequently approach expert performance in the last repetition of each phrase. 
We average the typing speed of the third repetition of each phrase, and found that the typing speeds in this setting increased $12$-$45\%$, with final typing speeds of $16.4$-$27.8$ wpm (average: $23.3$~wpm, Table \ref{table:typing_speed}). 
Compared to LowKeyEMG-base, word completion, context, and expertise combine to yield a total $42$-$112\%$ wpm improvement.

\begin{table}[b!]
  \centering
\begin{tabular}{
    >{\raggedright\arraybackslash}m{2.0cm}
    >{\raggedright\arraybackslash}m{1.8cm}
    >{\raggedleft\arraybackslash}m{0.9cm} >{\raggedright\arraybackslash}m{1.8cm}
    >{\raggedleft\arraybackslash}m{0.9cm} >{\raggedright\arraybackslash}m{1.7cm}
    >{\raggedright\arraybackslash}m{1.7cm}
  }
    \toprule
Participant     & \multicolumn{1}{c}{\thead{LowKeyEMG\\base}}&& \multicolumn{1}{c}{\thead{LowKeyEMG\\completion}}&& \multicolumn{2}{c}{\thead{LowKeyEMG\\completion+context}}\\
    & & & & &\thead[l]{interleaved}&\thead[l]{isolated} \\
\toprule
H1 & $19.6\pm4.0$ \hspace{0pt} & \scriptsize{$(+11\%\mathbin{=})$} & $21.7\pm4.0$ \hspace{4pt} & \scriptsize{$(+13\%\mathbin{=})$} & $24.4\pm4.0$ \hspace{0pt}& $27.8\pm6.3$\\
    H2 & $7.7 \pm3.6$ \hspace{0pt} & \scriptsize{$(+31\%\mathbin{=})$} & $10.1\pm3.6$ \hspace{4pt} & \scriptsize{$(+12\%\mathbin{=})$} & $11.3\pm3.6$ \hspace{0pt}& $16.4\pm3.4$\\
    H3 & $16.0\pm4.9$ \hspace{0pt} & \scriptsize{$(+25\%\mathbin{=})$} & $19.9\pm4.9$ \hspace{4pt} & \scriptsize{$(+14\%\mathbin{=})$} & $22.8\pm4.9$ \hspace{0pt}& $25.6\pm5.8$\\
    \midrule
    Average & $14.4$ wpm & \scriptsize{$(+19\%\mathbin{=})$} & $17.2$ wpm & \scriptsize{$(+13\%\mathbin{=})$} & $19.5$ wpm & $23.3$ wpm \\
 
    \bottomrule
  \end{tabular}
    \vspace{8pt}
    \captionof{table}{Typing Speed for 3 LowKeyEMG conditions, in words per minute (wpm, mean $\pm$ std).}
  \label{table:typing_speed}
\end{table}

\subsection{User experience}

We characterized user behavior, quantifying typing efficiency, online decoding accuracy and how users took advantage of LowKeyEMG's typing features. The efficiency of LowKeyEMG in closed loop use can be quantified by gestures per character (GPC), the ratio between the number of gestures used to type a sequence and the sequence length. 
To characterize the efficiency of each participant's usage of LowKeyEMG, we additionally subtract $2\cdot$\textit{undo} gestures from the total gestures of each phrase, thereby removing the influence of user gesture error or sEMG decoder error (errors are still shown for each user in Fig. \ref{fig:wpm}b). 
We found little variation in this error corrected GPC (blue bars, Fig. \ref{fig:wpm}b). Across participants, we observed an average GPC of 1.11, 0.95, and 0.83 GPC for LowKeyEMG-base, LowKeyEMG-completion, LowKeyEMG-completion+context, corresponding to efficiencies of $-11\%$, $5\%$, and $17\%$, respectively. The errors and undo gestures can come from several sources, such as user error, gesture detection false positive, or gesture misclassification. 
Participant error rates varied depending on user skill level and decoder quality, with ranges of $5.8$-$9.5\%$ (H1), $20.0$-$22.7\% $ (H2), and $10.4$-$13.9\%$ (H3).  
Participants typically chose to undo all errors without benefiting from distance-1 matches.

\begin{figure}

    \centering
\includegraphics[width=5.5in, trim = 0cm 0.0cm 0cm 0.0cm, clip]{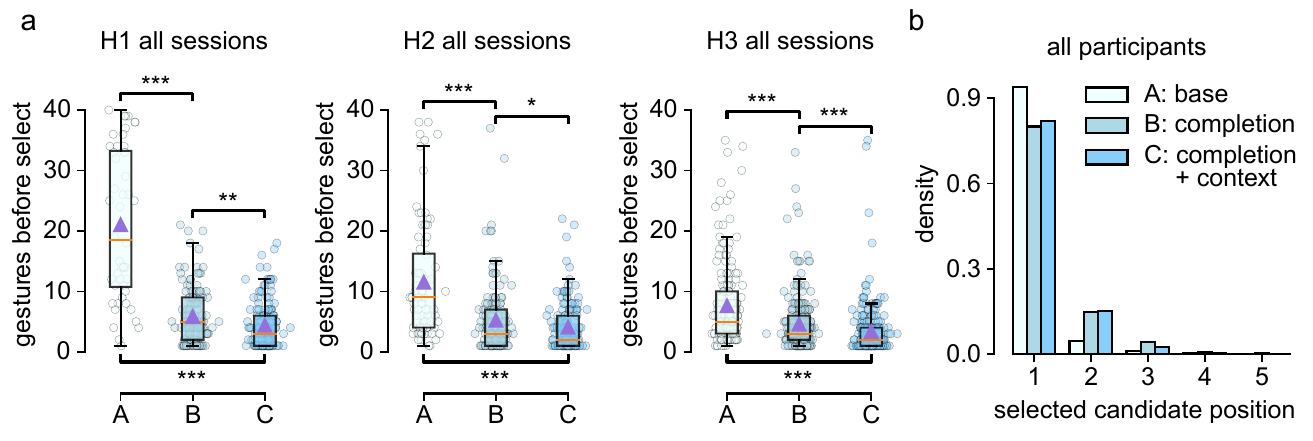}
    \caption{ 
    \textbf{Closed loop one-handed typing selection statistics.} 
    \textbf{(a)} The number of gestures before selection is shown for each participant for conditions A: base, B: completion, and C: completion+context.
    One-sided rank-sum test: $p<~$*$0.05$, **$0.01$, ***$0.001$. All participants select word suggestions significantly sooner with word completion (B), and when RWKV recieves additional passage context (C).
    \textbf{(b)} Probability distribution of position of selected candidate suggestions aggregated (equally weighted) across participants for conditions A: base, B: completion, and C: completion+context. When participants select a word, it is in the top-2 candidates at least 94.8\% of the time (condition B). For condition A and C, the selected candidate is in the top-2 97.3\% and 98.6\% of the time. 
    \label{fig:user_stats_selection}
}
\end{figure}

Participants realized higher efficiencies with LowKeyEMG across conditions by more frequently using word completions as they were enabled (condition B) and improved with passage context (condition C). 
Participants rarely selected suggestions outside of the top 2 across all conditions regardless of word completion and passage context (Fig. \ref{fig:user_stats_selection}b), with at least 94.8\% of the selections being in the top 2 for condition B, which had the least top 2 selections. 
On the other hand, gestures typed before selection of a suggestion decreased significantly for each user with the addition of word completions, and again when word completions were computed with passage context (Fig. \ref{fig:user_stats_selection}a).
User efficiency differences across conditions (Fig. \ref{fig:wpm}b) therefore emerged not in selection candidate depth at the time of selection, but in selecting \textit{earlier}.

\subsection{LowKeyEMG optimizations}

\begin{figure}
  \centering
\includegraphics[width=5.5in, trim = 0cm 0.0cm 0cm 0.0cm, clip]{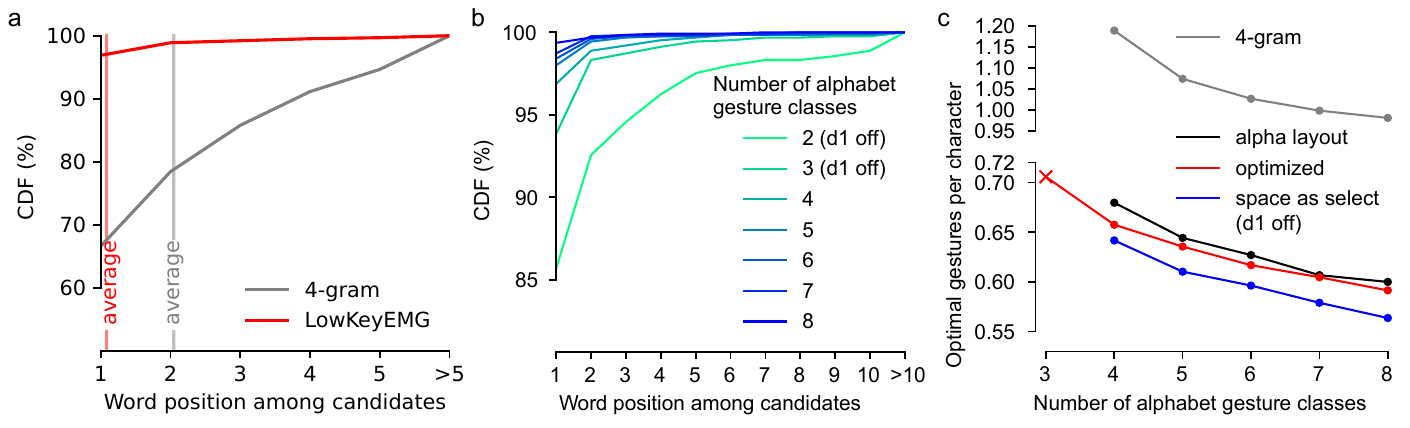}
  \caption{
  \textbf{Simulated results.}
  \textbf{(a)} Cumulative distribution functions of the position of each word among candidates after typing the entirety of each word and a space, using an optimized layout with 4 alphabetic classes, for LowKeyEMG and a 4-gram word LM.
  \textbf{(b)} Same as (a), but for optimized layouts of 2-8 alphabetic classes, using LowKeyEMG.
  \textbf{(c)} Gestures per character (GPC) for a simulated user who minimizes GPC for a 4-gram word LM, for LowKeyEMG with alphabetical and optimized layouts (distance-1 matching on except for 3 classes), and for LowKeyEMG where \textit{space} selects the top candidate once the next character is typed.
  }
\label{fig:sim}
\end{figure}

LowKeyEMG, using RWKV, improves top-1 ($96.9\%$ vs. $ 66.6\%$) and top-3 ($99.2\%$ vs. $85.8\%$) accuracy over a 4-gram word model \cite{vertanen2019arpa} using the same 7-gesture mapping for simulated typing of 6 passages of lengths 877, 1774, 381, 1064, 908, and 1585 characters, all sampled from the beginnings of books in the public domain (Figure \ref{fig:sim}a).
We find that decreasing the number of character classes from 8 to 4 only decreases top-1, top-2, and top-3 accuracy from $99.4\%$, $99.7\%$ and $99.8\%$ to $96.9\%$, $98.9\%$ and $99.2\%$ (Figure \ref{fig:sim}b).
In fact, as few as 2 alphabetic classes can be used while still achieving a top-10 accuracy of $98.9\% $ (\href{https://johannes-lee.github.io/lowkeyemg/#SupplementaryVideo2}{Supplementary Video 2}).

LowKeyEMG also allows a simulated user to type with fewer gestures, needing 34\% fewer (0.66) gestures per character (GPC, Figure \ref{fig:sim}c) than direct typing, compared to 1.19 GPC with the 4-gram model.
We attribute this difference to the increased context length incorporated by RWKV due to its recurrent architecture.
Optimizing the gesture layout improved the GPC compared to alphabetical layouts for all numbers of gesture classes between 3 and 8 (Figure \ref{fig:sim}c).
The GPC metric can be optimized by selecting the top candidate once \textit{space} and another character are typed, or even further optimized when \textit{select} queues the second-top candidate instead (0.54 GPC with 4 alphabetic keys), though these optimizations may make the system less intuitive for users.
Of note, the keyset can be further reduced by combining \textit{space} and \textit{select} into a single key, treating each first instance as \textit{space} and subsequent instances as \textit{select}, provided that users make selections after every word.
In addition to closed-loop decoding, LowKey typing can also be used to type entire sequences of text without intermittent selection and with minimal correction, achieving a $0.6\%$ top-1 CER with 4 alphabetic key and 1 space key when no incorrect keys are pressed (Supplementary Figure \ref{fig:sim_cer}).

\section{Discussion} \label{sec_discussion}

Our results demonstrate performant and efficient EMG-based text entry with 7 keys, and feasible entry down to 5 keys using LowKeyEMG, achieving typing speeds of $23.3$~wpm and efficient gestures per character rates.
As a metric, GPC is most useful at low gesture rates, since at increasing gestures per second, a user's GPC generally increases toward unity due to the cognitive load of processing candidate text \cite{quinn2016cost}.
LowKeyEMG also supports proficient typists who type quickly and accurately, yielding low CERs without requiring constant candidate selection.
Still, future research should continue to improve EMG decoding to support users of all skill levels.
We additionally emphasize that although this paper focuses on LowKeyEMG's efficacy when decoding EMG, LowKey typing systems with small keysets are not limited to EMG computer interfaces.
As EMG devices become more accessible to healthy and motor-impaired users, they require a reliable method of text entry, even when the number of decoded classes is limited.
LowKeyEMG provides a low-cost, low-latency, and low-key typing framework, with minimal reductions in performance.

\textbf{Limitations:}
As presented, LowKeyEMG applies the same language modeling for all users.
To further increase performance, a personalized layer (e.g., n-gram) can be added to the language model, adapting to each user's typing patterns.
Based on user preference, obscenities can be removed from the dictionary.
Personalizations may also include the use of made-up words, slang phrases, or grammatically incorrect but desirable sentences, which must be scored appropriately.
In order to type words outside of the dictionary or words that are unlikely given the context, future LowKey typing systems should offer finer-grained control of typing individual alphabetical characters. Other improvements should also be made to allow users to edit text freely, rather than sequentially, while efficiently accounting for both past and future context.

LowKeyEMG's word filter and character model did not account for varying error distributions, e.g. higher error rates when differentiating different gestures involving the same finger, instead assuming homoscedastic gesture errors of 4+ letter words, and requiring users to perform manual error correction with undo actions.
Fortunately, the character model can be readily replaced with one that does model error distributions, and can also potentially accommodate the omission of spaces (Algorithm \ref{alg_open_lm_beam}).
Additionally, when the distance-1 penalty was statically applied, we found cases where the character model was detrimental to performance, such that typing the word \textit{irony:54252} at the beginning of a passage was not possible, even though it could be typed when distance-1 words were disallowed.
Thus the character model should be changed to be context-dependent, applying stricter penalties at the beginning of passages when less information is available.

\begin{ack}
The authors would also like to thank Abhishek Mishra and Nevin Liang for their helpful discussions on this work.
This work was supported by National Institutes of Health (NIH) award numbers DP2NS122037 and R01NS121097.
Kao is the inventor of intellectual property owned by Stanford University that has been licensed to Blackrock Neurotech and Neuralink Corp.
Lee and Kao have a provisional patent application related to AI copilots for brain-computer interfaces owned by the Regents of the University of California. 
Lee, Kaasyap, Hadidi, Zhou, and Kao have a provisional patent application related to LowKey typing owned by the Regents of the University of California. 
Kao is a co-founder of Luke Health, on its Board of Directors, and has a financial interest in it.
\end{ack}

\medskip

{
\small
\bibliography{refs}

}

\newpage \appendix

\section{Technical Appendices and Supplementary Material}

\renewcommand{\figurename}{Supplementary Figure}
\renewcommand{\tablename}{Supplementary Table}
\setcounter{figure}{0}
\setcounter{table}{0}

Additional Supplementary Information files, including Supplementary Videos 1 and 2 and analysis code, are available at \url{https://doi.org/10.5281/zenodo.15588064}, and data is available at \url{https://doi.org/10.5281/zenodo.15492720}.

\begin{table}[H]
  \centering
  \begin{tabular}{lr}
    \toprule
    Parameter & Value \\
    \toprule
    Architecture & EEGNet \cite{lawhern2018eegnet} \\
    \# temporal filters & 64 \\
    \# spatial filters & 4 \\
    $p_{\mathrm{dropout}}$ & 0.5 \\
    Average pooling factor & 2 \\
    Loss function & Cross entropy \\
    Optimizer & Adam \\
    Learning rate & 0.001 \\
    Weight decay & 0.0001 \\
    $\epsilon$ & 0.001 \\
    Max epochs & 30 \\
    Patience of validation loss & 15 \\
    \bottomrule
  \end{tabular}
  \vspace{6pt}
  \captionof{table}{
  \textbf{EEGNet and optimizer hyperparameter settings.}
  The same hyperparameters were used for detection and classification.}
  \vspace{-0.7cm}
  \label{table:eegnet_hyper}
\end{table}

\begin{table}[H]
  \centering
    \begin{tabular}{
>{\raggedright\arraybackslash}m{2.0cm} >{\raggedright\arraybackslash}m{1.05cm} >{\raggedright\arraybackslash}m{1.05cm} >{\raggedright\arraybackslash}m{1.05cm} >{\raggedright\arraybackslash}m{1.05cm} >{\raggedright\arraybackslash}m{1.0cm} >{\raggedright\arraybackslash}m{1.0cm} >{\raggedright\arraybackslash}m{0.8cm} >{\raggedright\arraybackslash}m{0.8cm} }
    \toprule
Layout & Key I & II & III & IV & V & VI & VII & VIII \\
    \toprule
    2-optimized & {f g h i l n p r s t u v , : ; ! '} & {a b c d e j k m o q w x y z . ? -} & & & & & & \\
    \bottomrule
    3-optimized & {a h n r t w y : ; - '} & {b c e j k o s v z , !} & {d f g i l m p q u x . ?} & & & & & \\
    \bottomrule
    4-optimized (real-time experiments) & {a c d u v w x : ;} & {g j k l m o p y -} & {b e r t z , '} & {f h i n q s . ! ?} & & & & \\
    \bottomrule
    4-alphabetical & {a b c d e f g : ;} & {h i j k l m n o -} & {p q r s t , '} & {u v w x y z . ? !} & & & & \\
    \bottomrule
    5-optimized & {b d j k l o p q x z} & {h n u y : ;} & {a f t , - '} & {g i r s . ?} & {c e m v w !} & & & \\
    \bottomrule
    5-alphabetical & {a b c d e : ;} & {f g h i j -} & {k l m n o '} & {p q r s t , !} & {u v w x y z . ?} & & & \\
    \bottomrule
    6-optimized & {a b k s :} & {d h m u x ;} & {o t z -} & {e i v w ' !} & {c j l p q r ,} & {f g n y . ?} & & \\
    \bottomrule
    6-alphabetical & {a b c d e :} & {f g h i j ;} & {k l m n -} & {o p q r '} & {s t u v , !} & {w x y z . !} & & \\
    \bottomrule
    7-optimized & {e j s :} & {k l o q w y z} & {b h i x ;} & {c d p u v -} & {r t , '} & {a m . ?} & {f g n !} & \\
    \bottomrule
    7-alphabetical & {a b c :} & {d e f ;} & {g h i j -} & {k l m n '} & {o p q r !} & {s t u v ,} & {w x y z . ?} & \\
    \bottomrule
    8-optimized & {r s :} & {b k o q y ;} & {c e j p -} & {d i m '} & {l u v w x} & {h n . ?} & {f t z ,} & {a g !} \\
    \bottomrule
    8-alphabetical & {a b c :} & {d e f ;} & {g h i -} & {j k l '} & {m n o . ?} & {p q r ,} & {s t u v !} & {w x y z} \\
    \bottomrule
  \end{tabular}
  \vspace{6pt}
  \captionof{table}{
  \textbf{Key-character mappings for layouts with 2-8 alphabet keys.} Real-time experiments used the ``4-optimized'' layout, and did not use the characters \{: ; ! - '\}.}
  \vspace{-0.5cm}
  \label{table:layouts}
\end{table}

\begin{figure}[H]

    \centering
\includegraphics[height=2.8in, trim = 0.0cm 0.0cm 0cm 0.0cm, clip]{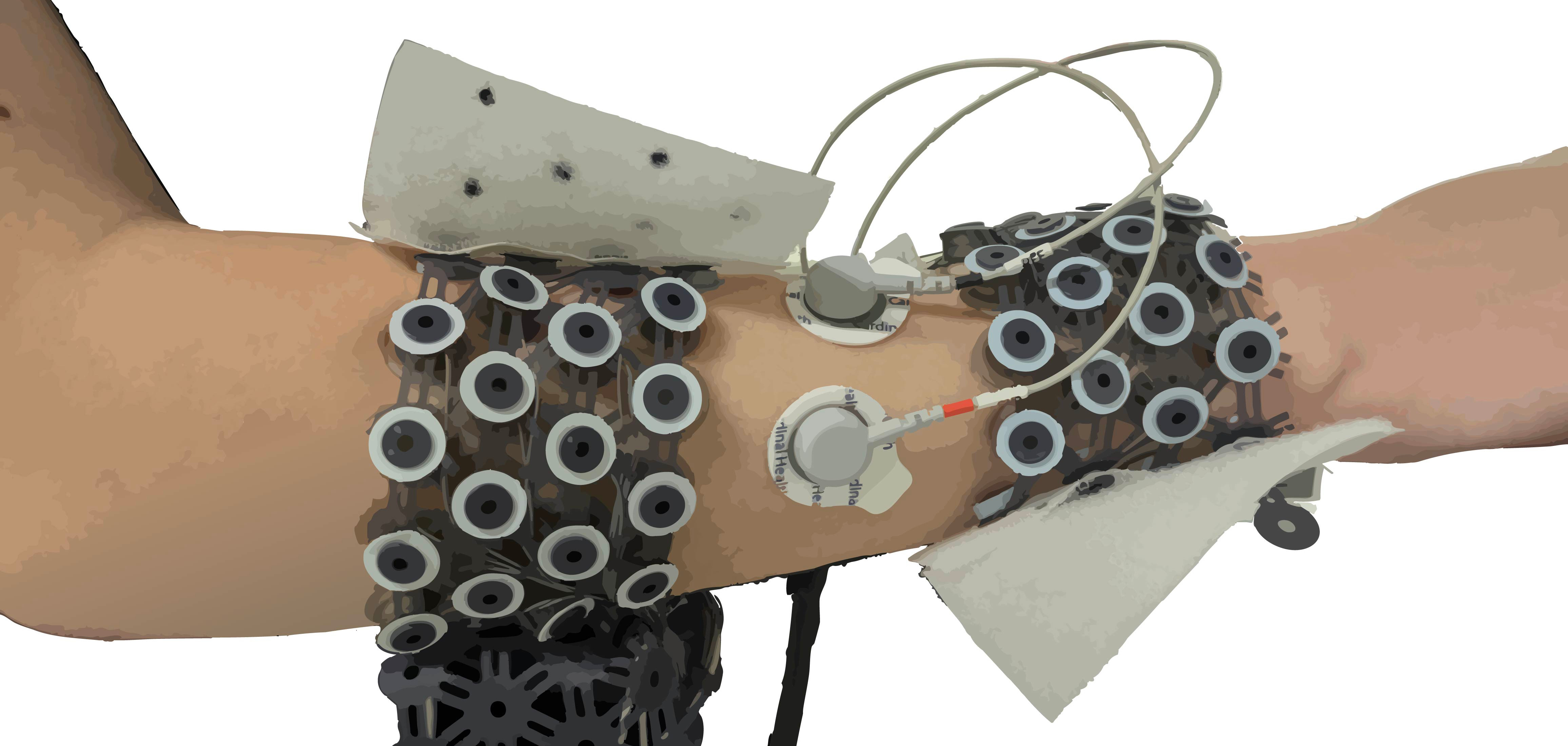}
    \caption{
    \textbf{Electrode placement of the two electrode arrays.} Explicit consent was acquired to release this image.
    \label{fig:electrode_placement}
}
\end{figure}

\begin{figure}[H]

    \centering
    \includegraphics[height=5.0in, trim = 0cm 0.0cm 0cm 0.0cm, clip]{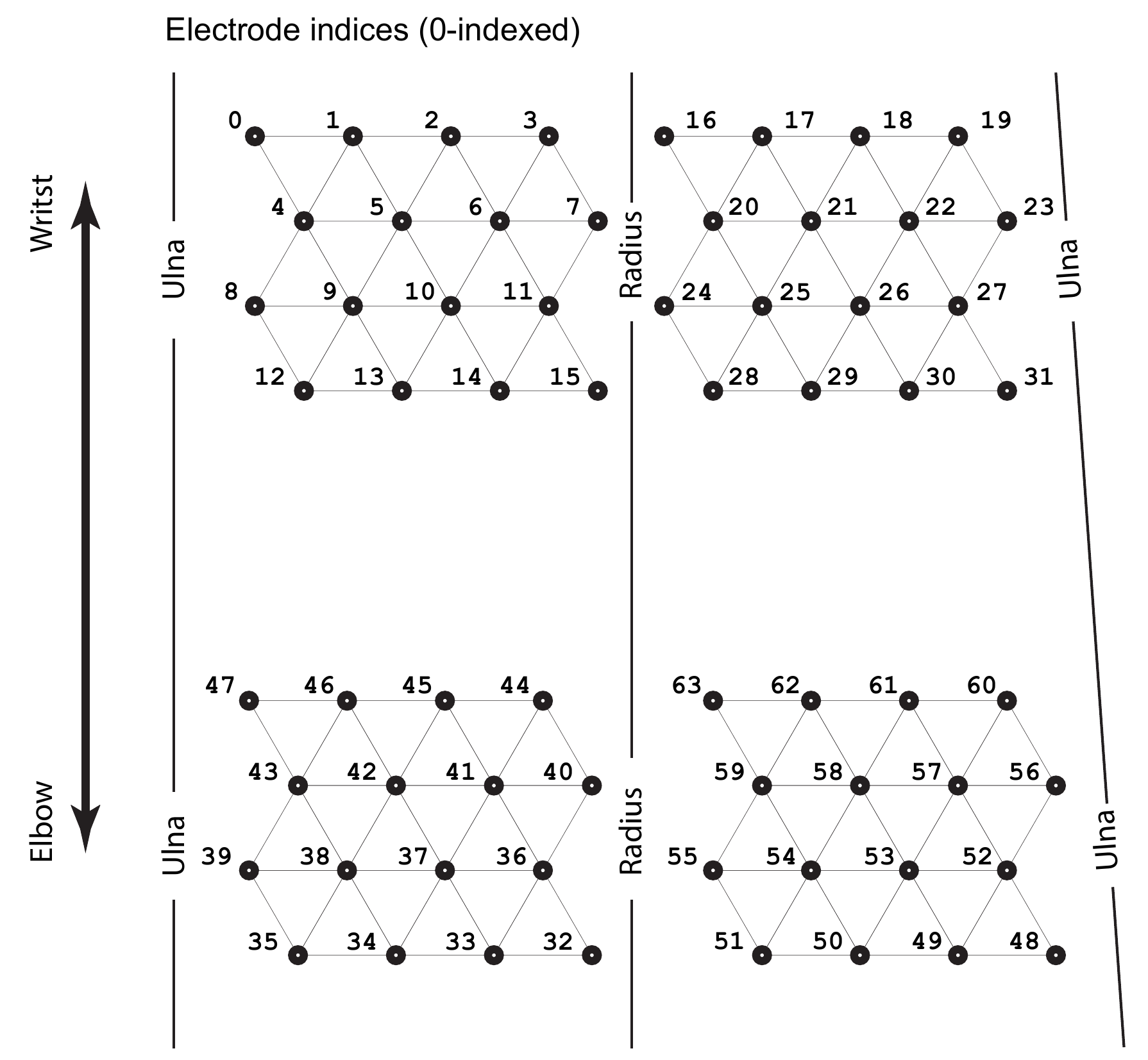}
    \caption{ 
    \textbf{Placement of electrodes and their corresponding 0-index.}
    \label{fig:electrodeidx}
}
\end{figure}

\begin{figure}[H]

    \centering
    \includegraphics[width=5.0in, trim = 0cm 0.0cm 0cm 0.0cm, clip]{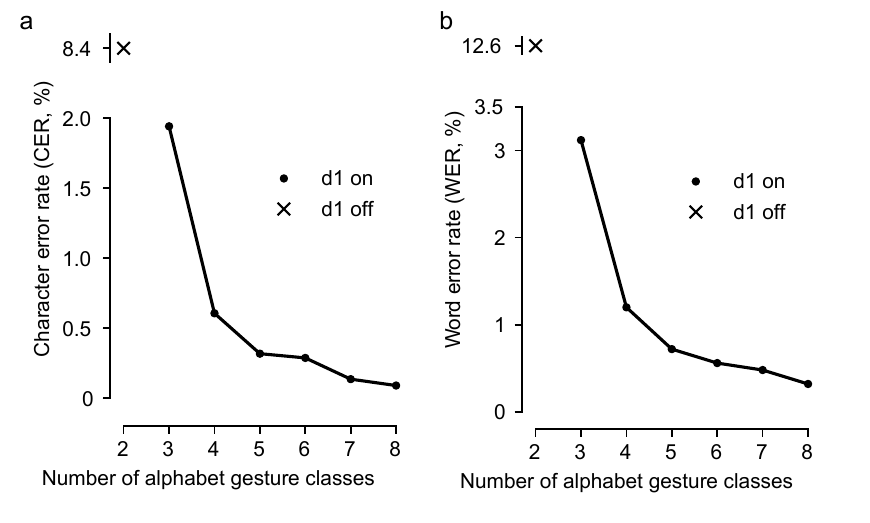}
    \caption{ 
    \textbf{Character error rate and word error rate without selection.} 
    \textbf{(a)} Character error rate (CER) when typing 6 English passages using varying alphabet gesture classes without selection, and with no key errors.
    \textbf{(b)} Same as (a), but for word error rate. Punctuation is treated as 1 word, as is the string \textit{'s}.
    \label{fig:sim_cer}
}
\end{figure}

\begin{figure}[H]

    \centering
    \includegraphics[height=2.7in, trim = 0cm 0.0cm 0cm 0.0cm, clip]{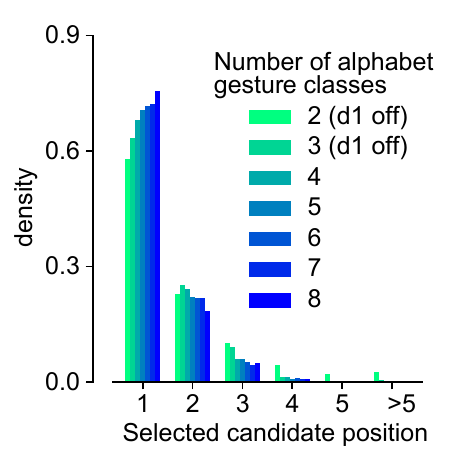}
    \caption{ 
    \textbf{Candidate selection positions when selecting every word in simulation.}
    Probability distribution of position of selected candidate suggestions for a simulated typist. This typist makes a selection for every word, and minimizes the gestures per character (GPC) while doing so.
    \label{fig:sim_sel_pos}
}
\end{figure}

\textbf{\href{https://johannes-lee.github.io/lowkeyemg/\#SupplementaryVideo1}{Supplementary Video 1}}:
Participant H1 performs the typing task using LowKeyEMG with 4 alphabetic keys, 1 space key, 1 select key, and 1 undo key.
Each of 7 one-handed gestures is mapped to 1 key.
H1 types 3 repetitions of a phrase under conditions C:~completion+context, A:~base, and B:~completion, respectively.

\textbf{\href{https://johannes-lee.github.io/lowkeyemg/\#SupplementaryVideo2}{Supplementary Video 2}}:
A user types phrases using 2 alphabetic keys, 1 space key, 1 select key, and 1 undo key, in direct input mode.
The context is reset to ``\texttt{<|endoftext|>}'' after every phrase.
Word completion is enabled and distance-1 matching is off.

\newpage

\end{document}